# Visualizing Collective Idea Generation and Innovation Processes in Social Networks

Yiding Cao, Yingjun Dong, Minjun Kim, Neil G. MacLaren, Sriniwas Pandey, Shelley D. Dionne,
Francis J. Yammarino, and Hiroki Sayama*
Binghamton University, State University of New York, Binghamton, NY13902, USA
ycao20@binghamton.edu; sayama@binghamotn.edu

**Abstract**

Collective idea generation and innovation processes are complex and dynamic, involving a large amount of qualitative narrative information that is difficult to monitor, analyze, and visualize using traditional methods. In this study, we developed three new visualization methods for collective idea generation and innovation processes and applied them to data from online social network experiments. The first visualization is the Idea Cloud, which helps monitor collective idea posting activity and intuitively tracks idea clustering and transition. The second visualization is the Idea Geography, which helps understand how the idea space and its utility landscape are structured and how collaboration was performed in that space. The third visualization is the Idea Network, which connects idea dynamics with the social structure of the people who generated them, displaying how social influence among neighbors may have affected collaborative activities and where innovative ideas arose and spread in the social network.

## 1 Introduction

Organizations are increasingly relying on collectives: groups of individuals which share expertise and take collaborative actions [1]–[4]. In such collectives, multiple people work together interdependently to achieve a greater goal than would be possible for individuals to accomplish alone [5]–[8]. Sharing of expertise among collective members with diverse backgrounds and behaviors is recognized as an important factor in collective effectiveness [9]–[12]. Accordingly, there is an increasing need for researchers to better understand the role that collaboration plays in collective design outcomes [13]–[15]. However, it is difficult to investigate the collective collaboration process in depth for several reasons [16], [17]. First, a collective is a higher level entity that can be composed of groups of groups, teams of teams, departments, or even organizations [18]: larger size and more complicated connections make investigation of a collective much harder than regular groups or teams [19]. Second, collective collaboration is generally a large-scale search and design processes typically involving heterogeneous individuals with diverse knowledge, expertise, and behaviors: a large number of variables could conceivably influence design outcomes [20]. Third, such collaborative tasks are usually knotty and open-ended with no simple solutions immediately available to anyone in the collective [19]. Consequently, the ideas produced by individuals in the collective are diverse and changing over time, making the collaboration process difficult to monitor. Furthermore, collective collaboration is a dynamic process that can take place over several days, weeks, months, or even years [22]–[24].

Another difficulty in examining how collaboration impacts collective outcomes is in how to acquire measurements with which to evaluate performance at multiple levels. Earlier human-subject studies typically evaluated performance only at one level using straightforward metrics such as the number of ideas generated [25], speaking time [26], performance scores [27], and win rates [28]. However, in most realworld scenarios, performance needs to be evaluated using multilevel measurements which examine the performance at both the individual and collective levels [29]–[31]. For example, when evaluating group work, people are not only concerned with what the whole group achieves but also with which group members made (or did not make) significant contributions [32].

When individuals of a collective come together to collaborate for problem solving, they often begin with a process of idea diversification, coming up with as many potential solutions as possible [33]. After that, the collective needs to narrow down the ideas to achieve convergence and obtain the best or most innovative solution [34], [35]. This process can be called collective idea generation and innovation. Many existing studies on collective idea generation and innovation attempt to detect idea divergence and convergence as a way of better understanding critical collective collaboration processes [36]; other studies focus on capturing the emergence of innovations [37]. Innovation is increasingly recognized as the leading goal of collective collaborative design, yet it is usually difficult to detect and track [38]. Furthermore, innovation is typically treated as a final achievement in a collective collaboration process, but it actually can occur at any time during the process [39]. Thus, it is essential to develop a way to detect novel ideas or solutions at any time during a collaboration and to identify which individual produced the innovative idea [37]. Finally, all of the divergence, convergence, and innovation processes include generating and selecting ideas. One of the most difficult problems in studying collective idea generation and innovation is that the ideas acquired from the records of an experiment are mostly written in a natural language which is difficult for a computer to understand and analyze using traditional quantitative analysis methods [40], [41]. In sum, collective collaboration processes are complex and dynamic, involving a large amount of qualitative narrative information that is difficult to monitor and analyze using traditional methods.

Visualization has been adopted as an effective way to obtain insights from complex, dynamic, and narrative data [42]–[44]. To learn about people's behaviors from social data, one needs to have a proper way to identify and analyze the dynamic information diffusion process. Efforts have been made to visualize social media data from multiple perspectives. For example, Marcus et al. [45] designed a timeline-based plot to a collection of Twitter data to display the dynamic behavior and interactions among Twitter users. Similarly, Dörk et al. [46] presented a timeline visualization to visualize a continuously updating information stream using Twitter online discussions.

Another aspect of research relates to social data visualization involving textual data and sentiment analysis using visual analytics [47], [48]. Zhang et al. [49] examined and extended a sentiment visualization method to visualize the opinion distributions of different Twitter groups. Liu et al. [50] designed a sedimentation-based visualization to enable the interactive analysis of streaming text data and their corresponding topical evolution. To sum up, visualization presents data and analysis results intuitively and consequently supports data interpretation and results validation. Moreover, successful visualizations enable users to obtain meaningful and high-level information from an overview as well as to ascertain the details of each data points.

In this study, we present three new visualization methods along with an online social network experiment as an example to demonstrate the functionality of these methods. The experiment provided a concrete scenario in which the key issues and challenges involved in collective activity and performance visualization can be considered and demonstrated. With the help of proposed visualization methods, users can monitor the collective idea generation activity and track idea clustering, transition, and innovation intuitively. The methods can also visualize narrative ideas with how the idea space is structured and how the collective collaboration performed in that space. Moreover, with the visualizations that connects idea dynamics with social structure of people who generated them, users can observe how social influence among neighbors may have affected collaborative activities and detect where innovative ideas arose and spread in the social network.

The potential users of these methods can be class instructors, project investigators, group managers, and organization consultants who employ the visualizations to accurately and efficiently evaluate individual/group performance and outcomes on a collaboration project. These methods can also help group managers or project designers to better understand the collective dynamics and thus appropriately adjust various collaboration parameters to further improve idea generation and innovation.

## 2  Online social experiment

We demonstrated how the proposed visualization methods can be applied to collective idea generation and innovation processes using the data obtained from an online social network experiment we conducted. In this experiment, participants were requested to develop a timeline of any product following the format of the examples given in the task description (Table I). All the communication and collaboration among the participants took place on a custom-built online social media-like platform we built in-house using Python and Flask and hosted in our laboratory's web server [21]. In this task, all the ideas generated by the participants were open-ended free-form narratives. The participants in this experiment were undergraduate/graduate students at a mid-size U.S. public university who were majoring in various disciplines, including Management (e.g., Accounting, Management Information Systems, Marketing), Engineering (e.g., Industrial and Systems Engineering, Systems Science), and other disciplines (e.g., Psychology, Actuarial Science, and Mathematical Sciences). They were assigned to collectives with 25–26 members each, connected in a predetermined topology of a ring-shaped regular network in which each node had four neighbors [51]. These network connections did not change during each experiment. Each collective had diverse individuals with multiple academic backgrounds.

TABLE I
Online experiment task instruction

| |
|---|
| *Design Task:* Presented below are two examples of the developmental timeline of products: |
| *Example 1: Personal communication device*<br>*Telegraph → land line → mobile phone → smart phone → ???*<br><br>*Example 2: Calculator*<br>*Abacus → mechanical calculator → electronic calculator → ???* |
| *In a similar way, you are to predict/design a future state for another product. Please describe **which product will undergo what kind of technological transition** (include only one transition in your idea).*<br><br>*This is a group activity, so you should collaborate with others by liking and commenting on other ideas shown below.* |

One online experimental session was two weeks long, during which participants were requested to log in to the experimental platform using anonymized usernames. Participants were requested to spend at least 15 min each weekday working on the assigned collective design task through collaboration with their anonymous social neighbors on the online platform: posting novel or modified ideas to the platform and discussing by reading, liking, and commenting on ideas posted by their neighbors. By utilizing their neighbors' ideas and responses/comments received from them, participants were expected to continuously elaborate and improve their idea quality over time. After the two-week session was over, participants received a final reminder email which asked them to complete an end-of-session survey form within one week. This survey asked participants to provide three final ideas they thought were the best among the posted ideas available on the platform. These final ideas were later evaluated by thirdparty experts who had expertise in product design and did not participate in the experiment or know any information about our experimental design or protocol. These evaluation results were used to quantitatively assess the quality of the final ideas developed by each collective.

The experimental settings simulated, to a limited extent, a real-world collective idea generation and innovation process. A collective in our experiment was large and involved people with diverse knowledge, expertise, backgrounds, and behaviors. Like a typical real-world collective collaboration process, the duration of this experiment was relatively long. Additionally, the design task used in this experiment was open-ended and challenging for participants, stimulating them to display exploration and exploitation behaviors, and making the collaboration process more complex. Participants' ideas changed over time, and the discussion record contained a large amount of qualitative narrative information which would be difficult to analyze using traditional methods.

The three new visualization methods proposed in this article are developed to address the difficulties mentioned above. These methods also allow for evaluating performance of the whole collective as well as that of each individual member.

## 3 Data Preprocessing

Using the data obtained from the experiment described in Section II, here we describe how the three visualizations are developed, and also identify issues that need to be considered when users develop visualizations using their own data.

For our three visualization methods, some data preprocessing is needed. As mentioned earlier, the idea data acquired from the experiment were mostly in free-form text format. Therefore, we preprocess the text data using a semantic embedding algorithm to convert them into numerical representations. We call this step idea embedding. There are many semantic embedding algorithms, such as bag of words (BOWs) [52], Latent Dirichlet Allocation (LDA) [53], word2vec [54], and doc2vec [55]; we used doc2vec in this study. Doc2vec, an adaptation of word2vec, is an unsupervised machine learning algorithm that can generate numerical vectors as a representation of sentences, paragraphs, or documents. Compared to other algorithms, doc2vec can provide a better text representation with a lower prediction error rate because it can recognize the word ordering and semantics which are not addressed by other algorithms [56].

In our experiment, all daily ideas posted on the experimental platform and all final ideas submitted in the endof-session form were converted to numerical vectors using a single doc2vec model. We set the dimension of output vectors to 400 to allow the numerical vectors to represent as much information from the original idea as possible. However, many of the 400 dimensions in the vectors obtained with doc2vec were undoubtedly correlated with each other, making the dataset highly redundant. Therefore, we applied principal component analysis (PCA) to the set of idea vectors for each experimental session to reduce dimensionality [57]. We captured 58.52% and 30.21% of the data variance, respectively, with the first and the second principal components (PCs). Thus, the first two PCs, PC1 and PC2, were used to represent the idea vectors for the visualizations proposed in this article.

The data variance of the first two PCs is very important in deciding whether visualizations using those two dimensions are meaningful and useful. As mentioned earlier, PC1 and PC2 captured 58.52% and 30.21% of the total data variances in our example, respectively. This means nearly 90% of the total data variance was explained by the first two PCs. When people use our visualization methods, they should also check how much data variance is captured in PC1 and PC2 and if these values are reasonably high according to their data quality and other requirements of their own. Moreover, there are many other dimensionality reduction algorithms, such as multidimensional scaling (MDS) [58] and t-distributed stochastic neighbor embedding (t-SNE) [59], which might be more effective (and our proposed visualization methods work with any of them). People can try multiple dimensionality reduction algorithms to obtain the best visualization outcome which can convey more information about the idea features and their relationships.

## 4 Visualization Methods and Examples

Fig. 1 shows the workflow of how the proposed visualizations are developed. First, the collection of ideas generated from the whole duration of the experiment was used to build a doc2vec model that converted the ideas to 400-D numerical vectors. Like in any other unsupervised machine learning algorithms, one should ensure that the data size is large enough to build a reliable doc2vec model. In our example, we had about 350 ideas (free-form narratives) as the input to build the doc2vec model.

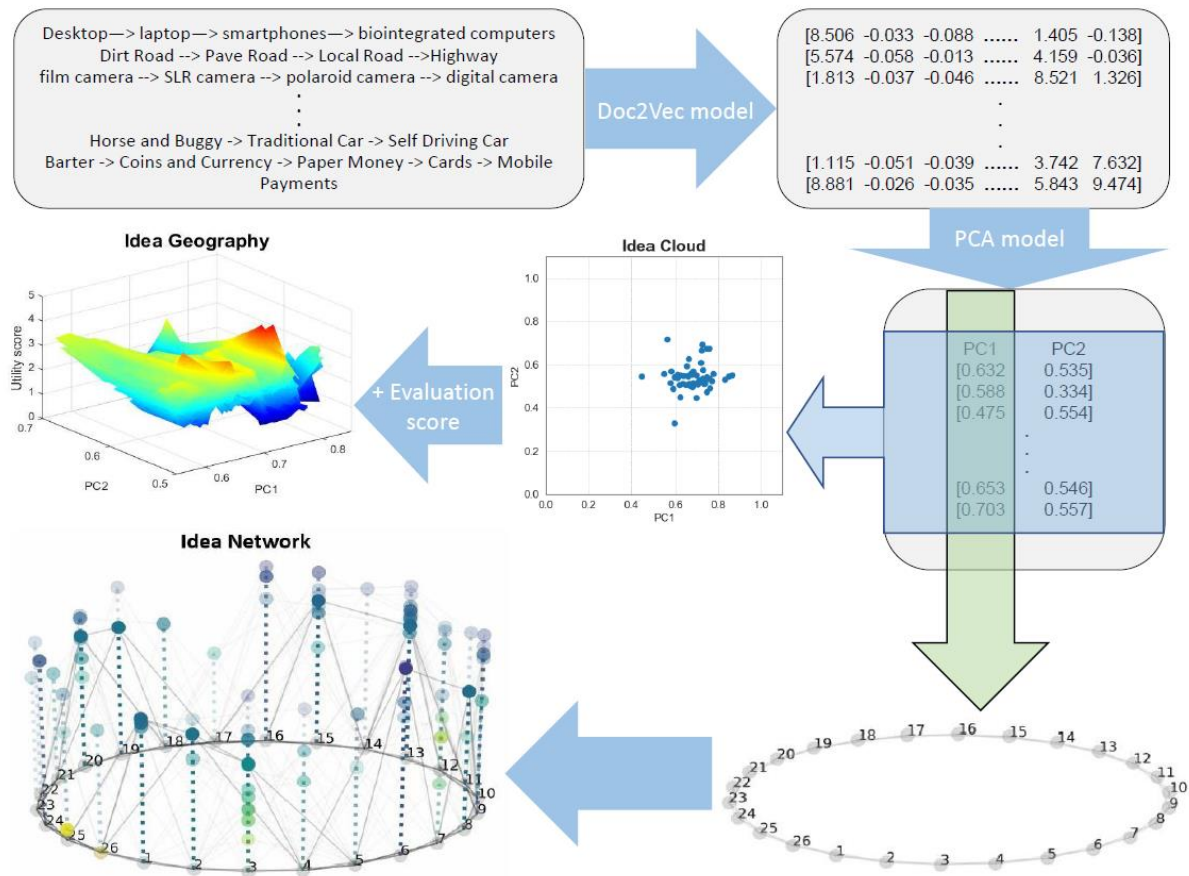

Figure 1 Overview of the workflow of proposed visualization methods. See text for details.

After the numeric idea vectors were obtained using the doc2vec model, PCA was performed on these vectors to obtain PC1 and PC2 values, which were used to plot an Idea Cloud. The 2-D problem space of an Idea Cloud provided the basis to construct an Idea Geography. Meanwhile, with the social network structure, the PC1 values were also used to build an Idea Network.

## 4.1 Idea Cloud

The first visualization method, which we call an "Idea Cloud," is a straightforward scatter plot of idea distributions as a point cloud in a 2-D space made of the first two PCs (PC1 and PC2). A scatter plot is a widely used visualization method for multiple purposes because of their relative simplicity, familiarity with users, and high visual clarity [60]–[62]. For example, the Idea Cloud in Fig. 2(a) was made using PCA. The 2-D PC space offers an efficient, intuitive way to monitor the locations of ideas and thereby check which ideas are relatively similar to or different from each other.

To check the validity of the dimensionality reduction method used in our visualization methods, we randomly selected ten ideas (three about audio devices, three about transportation, two about music players, and two about cameras) and obtained their locations in Fig. 2 (Table II). We find that similar ideas in the same category (highlighted in the same color) are also located closely in the PCA-based space and unsimilar ideas are generally distant from each other. This implies that PCA had a good performance in representing semantic distances in the 2-D visualization space in our example. We will therefore use PCA for dimensionality reduction and visualization in the rest of the article. Meanwhile, other dimensionality reduction methods may be used instead, such as t-SNE [shown in Fig. 2(b)] that recovers local semantic distances well.

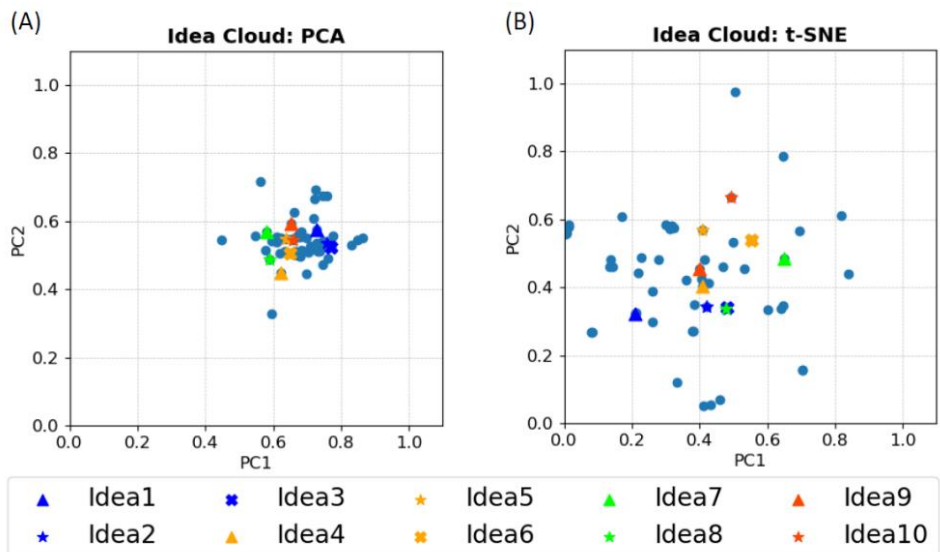

Figure 2 Idea Cloud examples (A) Using PCA (B) Using t-SNE *Note: Sample ideas in TABLE II were labeled in this figure*

TABLE II
Sample ideas and idea locations shown in Figure 2

| No. | Idea Feature | Coordinates in (A) | | Coordinates in (B) | |
|---|---|---|---|---|---|
| | | PC1 | PC2 | PC1 | PC2 |
| 1 | Wired earbuds→Bluetooth earbuds→Wireless earbuds→micro-earbuds | 0.7284 | 0.5761 | 0.2109 | 0.3218 |
| 2 | over-ear headphones→earbuds→wireless earbuds→wireless earbuds control apps | 0.7593 | 0.5374 | 0.4191 | 0.3428 |
| 3 | Headphones→Earbuds with wires→earbuds without wires→Microchip | 0.7693 | 0.5237 | 0.4796 | 0.3391 |
| 4 | Steam Car→ Motor Car→ Hybrid Car→ Self-Driving Electric Car | 0.6227 | 0.4489 | 0.4072 | 0.4038 |
| 5 | Horse and Buggy→Traditional Car→Self Driving Car | 0.6388 | 0.5481 | 0.4087 | 0.5679 |
| 6 | Horse→Carriage→Bicycle→Car→Self Driving Car→Hovercar | 0.6490 | 0.5060 | 0.5541 | 0.5398 |
| 7 | MP3 player→mp4 player→mp5 player | 0.5796 | 0.5707 | 0.6502 | 0.4842 |
| 8 | Music Box → Boombox→ MP3→ Ipods/Wired Earphones→ Wireless Earphones | 0.5905 | 0.4884 | 0.4786 | 0.3372 |
| 9 | film camera→SLR camera→polaroid camera→digital camera→action cameras | 0.6528 | 0.5940 | 0.3995 | 0.4538 |
| 10 | Film Cameras→Digital Cameras→Internal Memory Recording | 0.6596 | 0.5445 | 0.4922 | 0.6650 |

Like some previous research that analyze social media data using a 2-D map [63], [64], the Idea Cloud can be used as a basic spatial map of ideas as shown in Fig. 3(a). Like a geographic map showing object locations, the Idea Cloud can display the idea point location in the idea space in a simple and visual way. For example, we can identify ideas which are located near, or far away from, the center of the idea cluster in the Idea Cloud, which may represent mainstream or unique ideas, respectively.

The Idea Cloud visualization method can also help us monitor how many key topics exist in the collective discussion. In the example shown in Fig. 3(b), a k-mean clustering algorithm was performed to the first two PCs, and the optimal number of clusters (k) was determined using the elbow method [65]. The idea points that belong to different clusters in Idea Cloud are visualized using different colors so one can see which ideas belong to which topic. We listed two randomly selected ideas for each cluster in Table III to show characteristic idea features in each cluster.

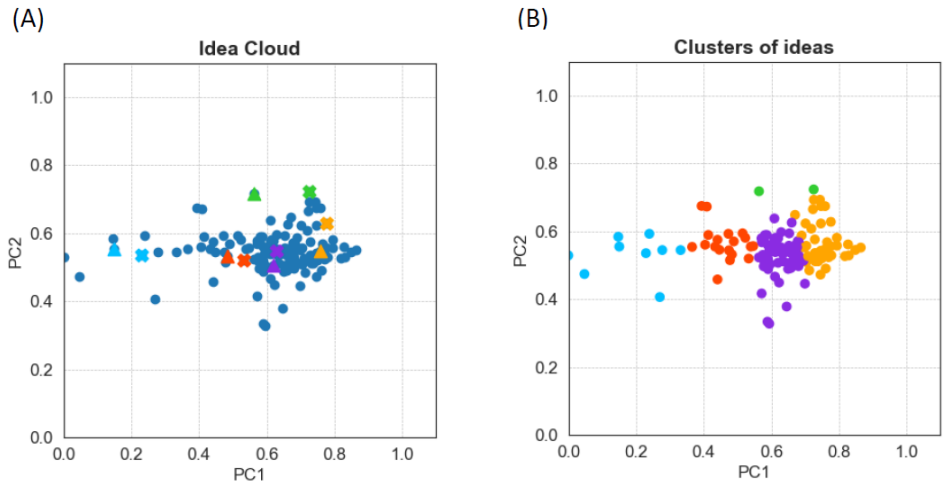

Figure 3 (A) Idea Cloud of whole collection of ideas. Note: Sample ideas in TABLE III were labeled in this figure
(B) Clusters of ideas

TABLE III
Sample ideas of each cluster in Figure 4

| Cluster No. | Idea No. | Idea |
|---|---|---|
| 1 | 1-1 | scrolls→books→audio books→Electronic readers |
| 1 | 1-2 | toothbrush→Electric Toothbrush→Self Brushing - toothbrush |
| 2 | 2-1 | Music Box→Boombox→MP3→Ipods→Iphone |
| 2 | 2-2 | Live Music→Record→Radio→CDs→iPod→Streaming→Virtual Musical Entertainment |
| 3 | 3-1 | Handwriting→Type Writing→Printing Press→Books→EBooks |
| 3 | 3-2 | Printing Press→Typewriter→Printer→3D Printer→Self assembling 4D printing |
| 4 | 4-1 | Mainframe→Mini-Computer→Workstation→Personal Computer→Laptop |
| 4 | 4-2 | Wired earbuds→Bluetooth earbuds→Wireless earbuds→micro-earbuds |
| 5 | 5-1 | merchants→stores→online shopping→drone delivery |
| 5 | 5-2 | lantern→bulb→tubelight→smart light bulb |

The Idea Cloud can be used to visually detect unique, potentially innovative ideas. For example, Fig. 4 compares the Idea Clouds of the same collective for two consecutive days, in which one can notice a new idea appeared on Day 2 [highlighted in red in Fig. 4(b)] in the area that was not explored on Day 1. This allows users to further inspect the original idea represented by this red dot to evaluate the innovativeness of this idea.

Idea Clouds are also useful for comparison of collective performance across multiple collectives. For example, Fig. 5 shows Idea Clouds of final designs submitted by two collectives under different experimental conditions. We can see that Collective 1 [Fig. 5(a)] generated a broader idea distribution than Collective 2 [Fig. 5(b)], which implies that Collective 1 may have produced more diverse ideas than Collective 2. Meanwhile, the average Euclidean distance between ideas (Avg_dis) for each Collective is presented in each plot that provides quantitative evidence for the idea diversity. Such a collective-level observation can help us better understand the relationships between conditions of collectives and their collective performance.

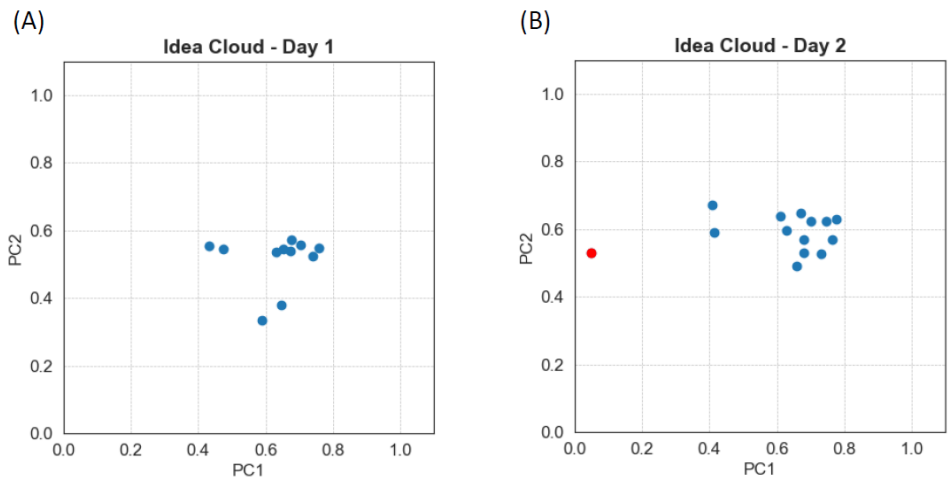

Figure 4 Idea Cloud of daily generated ideas (A) Idea Cloud of Day 1 (B) Idea Cloud of Day 2

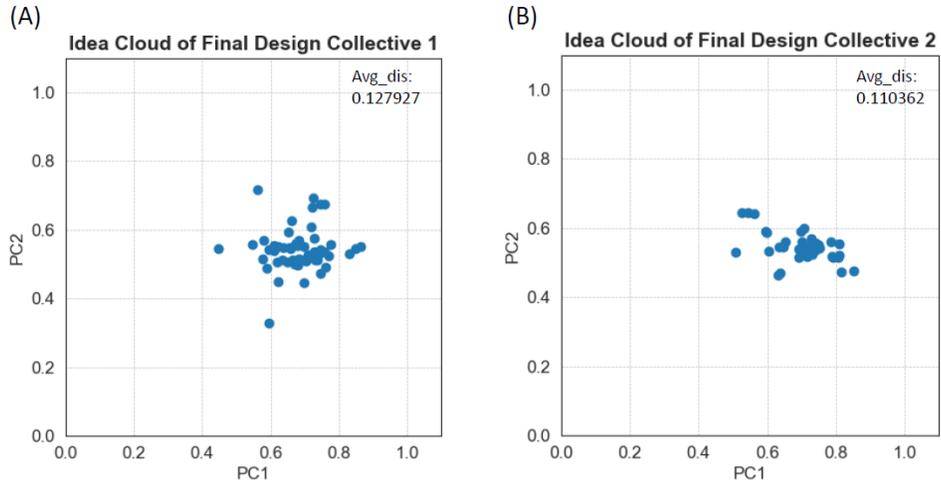

Figure 5 Idea Cloud of final designs (A) Idea Cloud of final designs of Collective 1 (B) Idea Cloud of final designs of Collective 2

Idea Cloud can also be used to track an individual's idea generation process over time, like GPS-based "run tracker" apps that are popular among runners who want to keep track of their running activities. Idea Cloud can track and visualize the sequence of ideas generated by one participant from the beginning to the end of an experimental session, with which one can evaluate the participant's exploration activities. For example, Fig. 6 shows such idea tracking results for two individual participants, where we can see that Participant #3 [Fig. 6(a)] had a longer "running" distance than Participant #13 [Fig. 6(b)], indicating that Participant #3 was more likely to have had a greater ability to explore new ideas and produce diverse ideas. The total distance recorded for each of them is also shown in the figure.

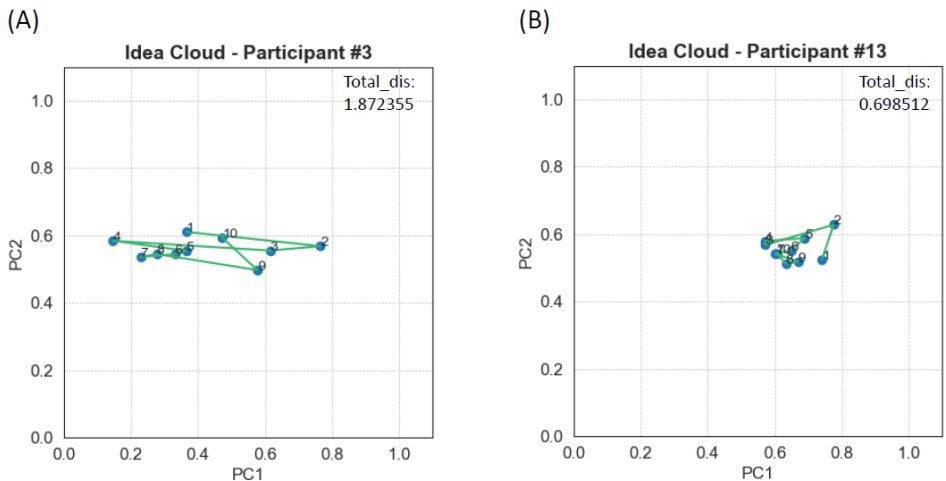

Figure 6 Idea Cloud of individual participants (A) Idea Cloud of Participant #3 (B) Idea Cloud of Participant #13

Moreover, by adding a third dimension for time, Idea Clouds can visualize temporal dynamics of collective idea generation. By combining multiple Idea Cloud plots along a time axis, we can intuitively trace how the idea distribution transitioned over time. Fig. 7 shows an example in which the Idea Clouds from. In this example, we can see that the idea distribution in Day 1 had a broader distribution, which means on the first day, participants produced more diverse ideas at the beginning of the session. On Day 2, the idea distribution became a little concentrated, meaning that the participants began to show some idea convergence due to the one-day information sharing and learning. Day 3 and Day 4 showed similar divergence and convergence patterns, and after Day 4, the idea distribution kept concentrated near the center until the end of the session. This kind of temporal visualization of idea distributions helps the experiment designer or group manager to better understand collective dynamics and thus determine various collaboration parameters, such as the appropriate length of a session.

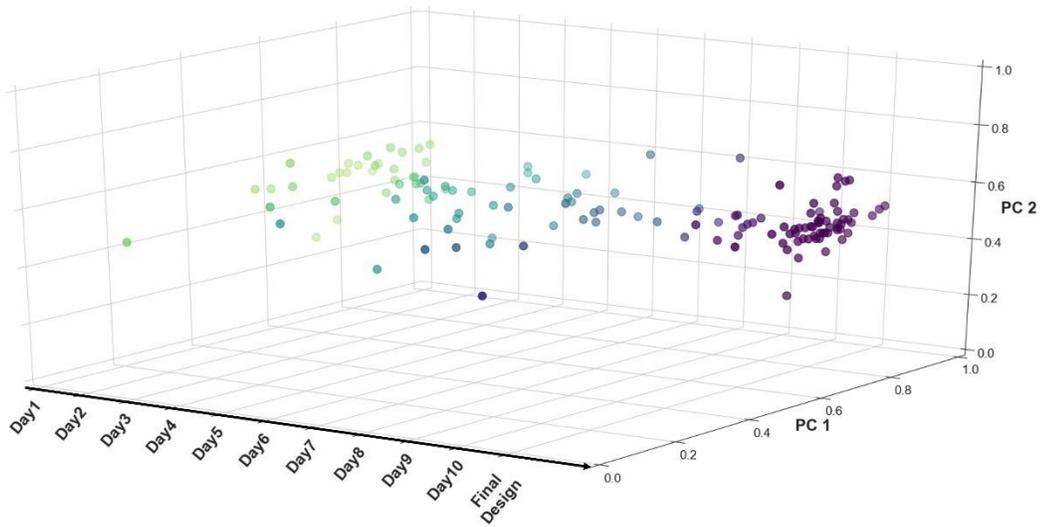

Figure 7 Time-series of Idea Clouds

## 4.2 Idea Geography

The second visualization method is what we call "Idea Geography." In an Idea Geography plot, the utility score which is represented by the mean value of evaluation score is used as elevation to construct utility terrains on the base of a 2-D Idea Cloud. Fig. 8 shows an example of Idea Geography, in which one can see mountains and valleys. The mountain areas represent regions populated by ideas with high evaluation scores, and on the contrary, the valley areas are populated by ideas with low evaluation scores.

Idea Geography can be useful to visualize and identify the location(s) of ideas which have the highest quality. For example, in Figure 8, the region with points "a" and "b" are easily identifiable as the "best idea" area. These two ideas described production timelines of headphones and earbuds, which were evaluated as the best final ideas of this collective.

Idea Geography can be used to compare the performance between multiple collectives. From the Idea Geography plots shown in Fig. 9, we can see that in Collective 2 [Fig. 9(b)] there is a clearly identifiable utility mountain area (to the left in the plot) where most of the submitted high quality final designs were concentrated. In contrast, Collective 1 [Fig. 9(a)] has a very small mountain area (to the right in the plot). This observation suggests that Collective 2 may have had greater ability to find the high utility area through the two-week idea exploration and exploitation period than Collective 1.

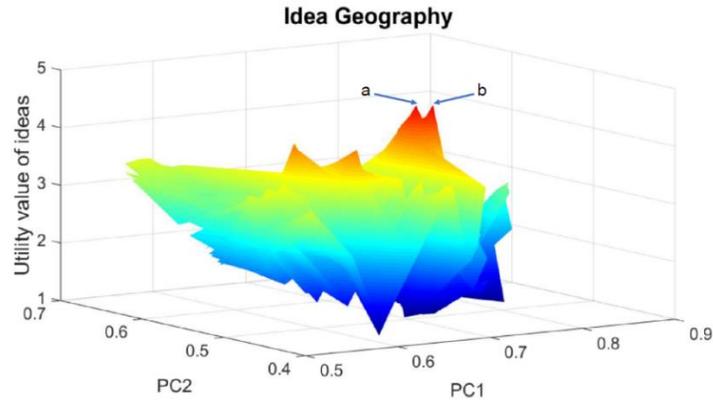

Figure 8 Idea Geography *Note: a and b represent two peaks in the Idea Geography*

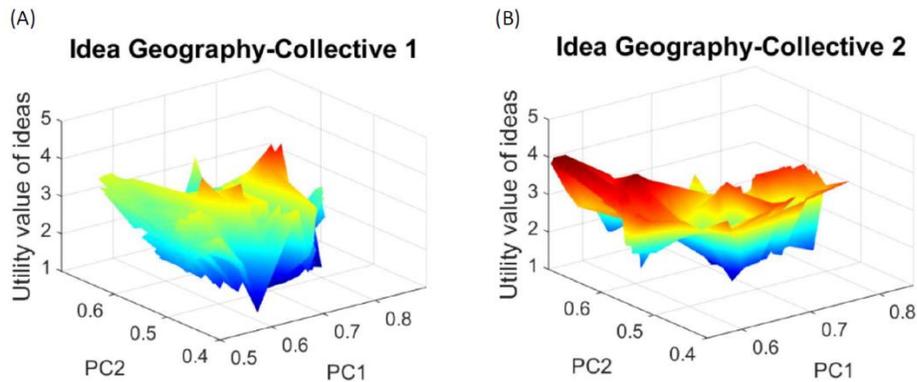

Figure 9 Idea Geography of Collectives (A) Idea Geography of Collective 1 (B) Idea Geography of Collective 2

Like Idea Cloud, Idea Geography can be used to evaluate not only the whole collective's performance but also an individual's performance as well. For example, in our experiment, participants were asked to submit three final designs. The idea points of final designs submitted by each participant can be marked in their collective's Idea Geography space with PC1, PC2, and utility value of these ideas. For example, in Fig. 10, we can see that Participant #4 made one final design which had the highest score, but the scores of the other two ideas were low. In comparison, Participant #21 may have had a better idea selection ability because, even though he/she did not make the best idea, Participant #21's idea scores were consistently high. Thus, Idea Geography can help us evaluate an individual participant's overall performance in the idea space.

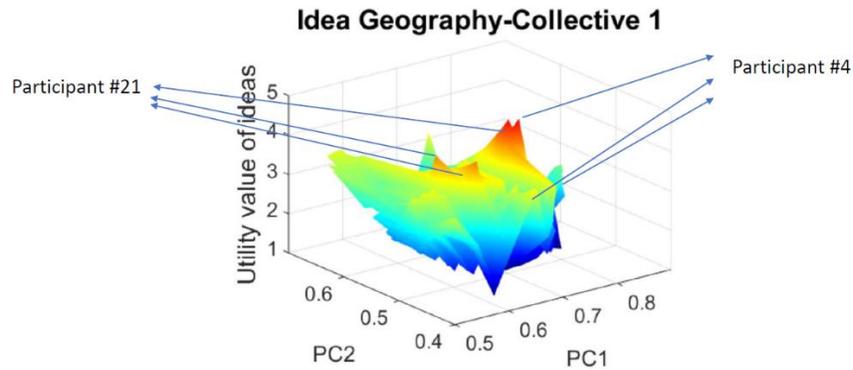

Figure 10 Idea Geography: Individual participants evaluation

The previous examples of Idea Geography used the evaluation scores of ideas for terrain elevation. We can also use elevation to represent other measurements according to what we need to examine. For example, Fig. 11 used the length of an idea as the elevation to estimate which ideas described a greater number of steps of future states of technological products. Comparing this Idea Geography with previous one in Fig. 8, we can find that the idea utility mountain area locates extremely close to the length of idea mountain area, which suggests that the longer, more elaborated ideas tended to receive higher evaluation scores.

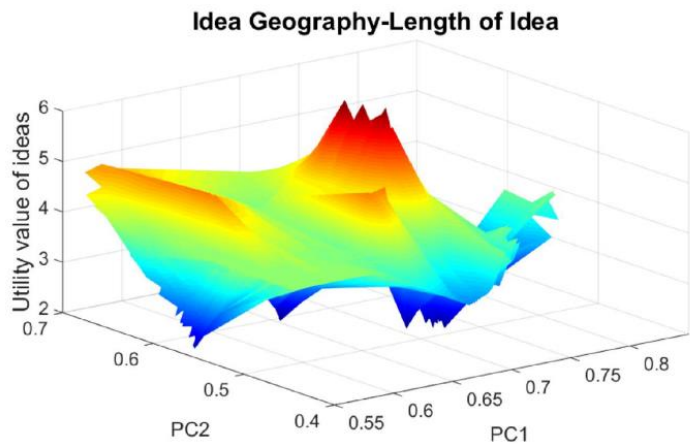

Figure 11 Idea Geography: lengths of ideas visualization

### 4.3 Idea Network

The third visualization method we propose is the "Idea Network." The Idea Network is a 3-D visualization constructed on a network based on traditional network layout which was made of human nodes and social connection links among them [66], [67]. In our experiment, the underlying social network was a spatially clustered regular network with node degree 4. More specifically, the participants were arranged in a shape of a ring, and each of them was connected to four nearest neighbors (two to the right and two to the left). Fig. 12 shows an example of an Idea Network. If a participant (a human node) submits an idea, an idea node appears above the human node. The height of the placement of the newly generated idea node is determined by the value of its first PC (PC1). The idea nodes are also connected to other idea nodes in the social neighborhood, following the topology of the social network. The shade of an idea node becomes less saturated as it becomes older to visualize the recency of ideas generated.

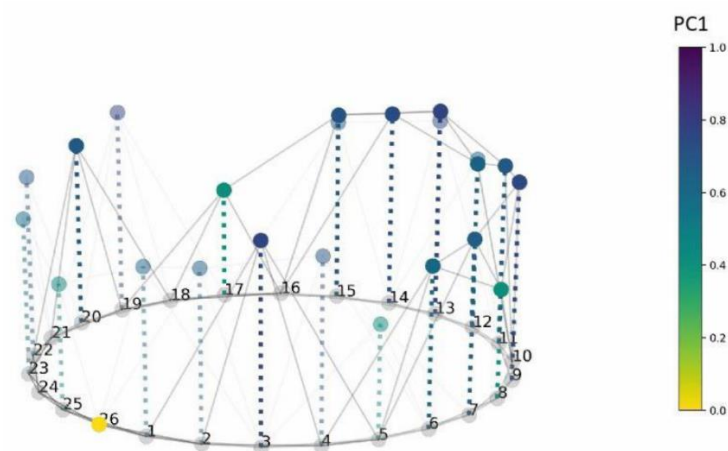

Figure 12 Idea Network

An Idea Network can be animated over time, connecting idea dynamics with the social structure of people who generated them and thus helping to observe how social influence among neighbors may have affected collaborative activities (an animation video can be found from link: http://shorturl.at/lBJ36). Users can easily rotate and zoomed-in view/out this interactive visualization to inspect details according to their specific needs. Such dynamic animations can offer an intuitive visualization of the idea spreading process in a social network.

Like the Idea Cloud that can detect the emergence of innovative ideas, the Idea Network can further show us where the innovative ideas arise. Fig. 13 shows an example. The yellow point which represents an idea submitted by Participant #26 had its PC1 value very different from that of the other ideas, indicating that Participant #26 produced a highly unique, potentially innovative idea. This helps capture the location of the innovation and detect which people are more innovative during collective collaboration.

The Idea Network can also evaluate an individual's idea exploration. We previously showed that the "run tracker" function of the Idea Cloud (Fig. 6) captured that Participant #3 had a high ability to produce diverse ideas. The Idea Network can provide the same answer in a social context. Fig. 14 shows that the ideas submitted by Participant #3 ranged widely in terms of the ideas' PC1 values, in stark contrast to the behaviors of her neighbors. This indicates that Participant #3's diverse ideas were not borrowed from his/her social neighborhood but were truly generated by Participant #3 as unique contributions to the collective collaboration. Besides, previous literatures indicated that people who speak a lot during the collective discussion may emerge as a leader [26], [68]. Therefore, this visualization may also help for detecting the emergence of leaders.

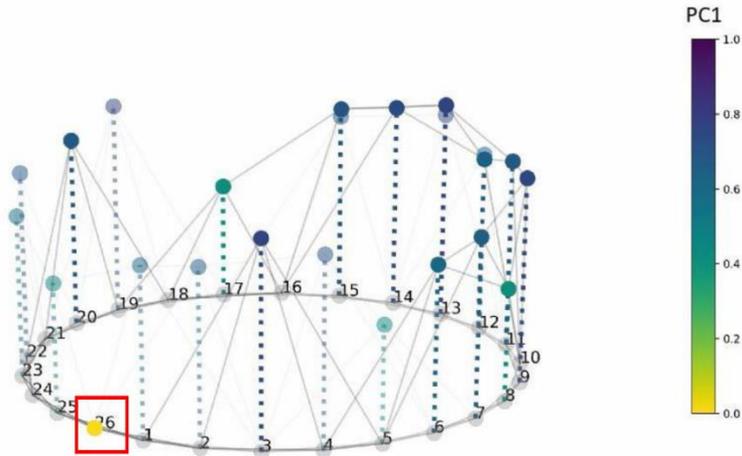

Figure 13 Idea Network: innovative ideas detection

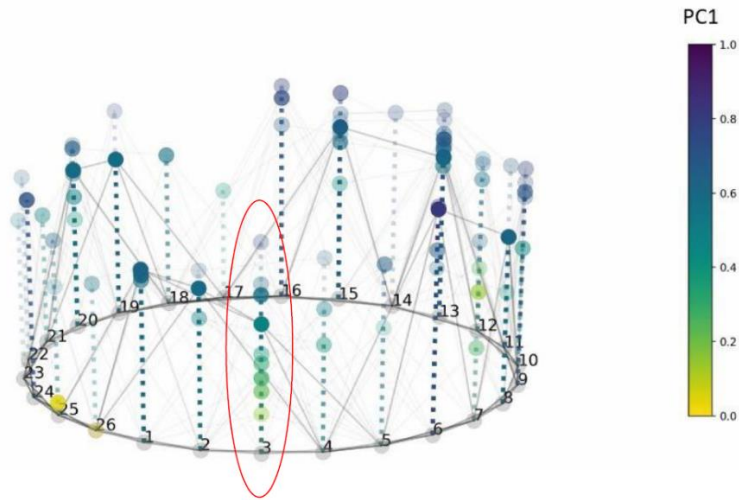

Figure 14 Idea Network: innovative individual detection

The last example of the use of Idea Network is evaluating collective-level efficiency. Fig. 15 presents an example. If we simply count the total number of ideas generated, we can see that Collective 2 generated more ideas than Collective 1. However, Fig. 15 also shows that, in Collective 2, there were several participants who made little to no contribution to idea generation (marked in red squares): most ideas were produced by a certain subset of people. Meanwhile, in Collective 1, almost every participant made contributions to the design and innovation process. This gives us an indication that Collective 2 might not be an efficient collective and it may be desirable to reorganize it or replace those noncontributing participants.

## 5 Conclusion

In this article, we proposed three methods for visualizing collective idea generation and innovation processes: Idea Cloud, Idea Geography, and Idea Network. The applications using our online experimental data represent a use case that exemplifies the benefits of these methods. In this use case, these visualizations provide easy, clear, and efficient ways to monitor and analyze collective activities and performance in the idea generation and innovation process. We expect these methods to be useful for studying complex collective collaboration processes which involve large-size collectives, open-ended tasks, and long durations.

The possible benefits of each of the proposed methods can be summarized as follows. The Idea Cloud can help an organization manager, or a collective member monitor collective activity and performance by tracking idea locations and the dynamic transitions of idea distributions over time during the whole collaboration process. Idea Clouds can also help detect clusters of key topics, detect the appearance of innovative ideas during discussion, and evaluate each individual participant's idea exploration ability. Idea Geography helps understand how the whole idea space is structured with regard to the ideas' utility values (or any other metrics of interest) and how the collective collaboration performed in the idea space; it can show the location(s) of the best ideas and help evaluate collective and individual participants' performance. Idea Networks are useful in visualizing idea dynamics in a social context, showing how social influence affects collaboration activity, and revealing where innovative ideas arise and how they spread in the network. Additionally, Idea Networks can provide information about each individual participant's contributions. These visualization methods offer many possible applications to researchers or human factors engineers in assessing collective performance over time, training interventions, and design features.

The three visualization methods excel at processing larger amount of ideas: in our case, hundreds to thousands of ideas generated by a collective in one experimental session. However, for a smaller number of ideas or small group idea generation analyses, limitations exist due to smaller sample size. To maintain the accuracy of idea vector prediction, the doc2vec algorithm typically needs a considerable amount of data. Additionally, all three visualization methods require enough data points to present a pattern of idea distribution or transition. A visualization method that can work well with both small and large datasets would be an important future development. Another limitation of the proposed visualization methods is that only when an experimental session is completed can we apply the methods to acquire the idea generation and innovation conditions for that session. Thus, it is very difficult for conducting periodic assessment during collaboration process, which is essential for incremental and adaptive adjustment for better collective configuration and performance. One worthy future work is to develop advanced Idea Cloud, Idea Geography, and Idea Network which can provide intermediate visualizations in the middle of a session, with which one can periodically check the collective collaboration progress and make appropriate adjustments as needed.

## Acknowledgments

This work was supported by the National Science Foundation under Grant 1734147

## Reference

[1] G. Stasser and S. Abele, "Collective choice, collaboration, and communication," Annu. Rev. Psychol., vol. 71, no. 1, pp. 589–612, Jan. 2020.
[2] R. A. Weber, "Managing growth to achieve efficient coordination in large groups," Amer. Econ. Rev., vol. 96, no. 1, pp. 114–126, Feb. 2006.
[3] S. Faraj and L. Sproull, "Coordinating expertise in software development teams," Manage. Sci., vol. 46, no. 12, pp. 1554–1568, 2000.

[4] M. Horton, P. Rogers, L. Austin, and M. McCormick, "Exploring the impact of face-to-face collaborative technology on group writing," J. Manage. Inf. Syst., vol. 8, no. 3, pp. 27–48, Dec. 1991.
[5] S. C.-Y. Lu, W. Elmaraghy, G. Schuh, and R. Wilhelm, "A scientific foundation of collaborative engineering," CIRP Ann., vol. 56, no. 2, pp. 605–634, 2007.
[6] J. Hammond, R. J. Koubek, and C. M. Harvey, "Distributed collaboration for engineering design: A review and reappraisal," Hum. Factors Ergonom. Manuf., vol. 11, no. 1, pp. 35–52, 2001.
[7] R. Sriram, S. Szykman, and D. Durham, "Special issue of collaborative engineering," J. Comput. Inf. Sci. Eng., vol. 6, no. 2, pp. 93–95, Jun. 2006.
[8] S. C.-Y. Lu and N. Jing, "A socio-technical negotiation approach for collaborative design in software engineering," Int. J. Collaborative Eng., vol. 1, no. 1, pp. 185–209, 2009.
[9] R. L. Moreland, J. M. Levine, and M. L. Wingert, "Creating the ideal group: Composition effects at work," in Understanding Group Behavior, 1st ed. London, U.K.: Psychology Press, 2018, pp. 11–35.
[10] D. van Knippenberg, C. K. W. de Dreu, and A. C. Homan, "Work group diversity and group performance: An integrative model and research agenda," J. Appl. Psychol., vol. 89, no. 6, pp. 1008–1022, Dec. 2004.
[11] H. van Dijk and M. L. van Engen, "A status perspective on the consequences of work group diversity," J. Occupational Organizational Psychol., vol. 86, no. 2, pp. 223–241, Jun. 2013.
[12] K. A. Jehn, G. B. Northcraft, and M. A. Neale, "Why differences make a difference: A field study of diversity, conflict, and performance in workgroups," Administ. Sci. Quart., vol. 44, no. 4, pp. 741–763, 1999.
[13] N. L. Kerr and R. S. Tindale, "Group performance and decision making," Annu. Rev. Psychol., vol. 55, no. 1, pp. 623–655, Feb. 2004.
[14] K. A. McHugh, F. J. Yammarino, S. D. Dionne, A. Serban, H. Sayama, and S. Chatterjee, "Collective decision making, leadership, and collective intelligence: Tests with agent-based simulations and a field study," Leadership Quart., vol. 27, no. 2, pp. 218–241, Apr. 2016.
[15] S. D. Dionne, J. Gooty, F. J. Yammarino, and H. Sayama, "Decision making in crisis: A multilevel model of the interplay between cognitions and emotions," Organizational Psychol. Rev., vol. 8, nos. 2–3, pp. 95–124, Oct. 2018.
[16] J. Shore, E. Bernstein, and D. Lazer, "Facts and figuring: An experimental investigation of network structure and performance in information and solution spaces," Org. Sci., vol. 26, no. 5, pp. 1432–1446, Oct. 2015. [17] W. Mason and D. J. Watts, "Collaborative learning in networks," Proc. Nat. Acad. Sci. USA, vol. 109, no. 3, pp. 764–769, Jan. 2012.
[18] F. Dansereau, F. J. Yammarino, and J. C. Kohles, "Multiple levels of analysis from a longitudinal perspective: Some implications for theory building," Acad. Manage. Rev., vol. 24, no. 2, pp. 346–357, Apr. 1999.
[19] K. L. Cullen-Lester and F. J. Yammarino, "Collective and network approaches to leadership," Leadership Quart., vol. 27, no. 2, pp. 173–180, Apr. 2016.
[20] R. J. Ely, "A field study of group diversity, participation in diversity education programs, and performance," J. Organizational Behav., vol. 25, no. 6, pp. 755–780, 2004.
[21] Y. Cao et al., "Capturing the production of innovative ideas: An online social network experiment and 'idea geography' visualization," in Proc. Int. Conf. Comput. Social Sci. Soc. Amer. Santa Fe, NM, USA, 2019, pp. 341–354.
[22] H. Sayama and S. D. Dionne, "Studying collective human decision making and creativity with evolutionary computation," Artif. Life, vol. 21, no. 3, pp. 379–393, Aug. 2015.
[23] J. A. Grand, M. T. Braun, G. Kuljanin, S. W. Kozlowski, and G. T. Chao, "The dynamics of team cognition: A process-oriented theory of knowledge emergence in teams," J. Appl. Psychol., vol. 101, no. 10, pp. 1353–1385, Oct. 2016.
[24] F. J. Yammarino, M. D. Mumford, M. S. Connelly, and S. D. Dionne, "Leadership and team dynamics for dangerous military contexts," Mil. Psychol., vol. 22, pp. S15–S41, Mar. 2010.
[25] K. Girotra, C. Terwiesch, and K. T. Ulrich, "Idea generation and the quality of the best idea," Manage. Sci., vol. 56, no. 4, pp. 591–605, Apr. 2010.

[26] N. G. MacLaren et al., "Testing the babble hypothesis: Speaking time predicts leader emergence in small groups," Leadership Quart., vol. 31, no. 5, Oct. 2020, Art. no. 101409.
[27] P. Kanawattanachai and Y. Yoo, "The impact of coordination on virtual team performance over time," MIS Quart., vol. 31, no. 4, pp. 783–808, Dec. 2007.
[28] A. Sapienza, Y. Zeng, A. Bessi, K. Lerman, and E. Ferrara, "Individual performance in team-based online games," Roy. Soc. Open Sci., vol. 5, no. 6, pp. 1–14, May 2018.
[29] S. D. Dionne and P. J. Dionne, "Levels-based leadership and hierarchical group decision optimization: A simulation," Leadership Quart., vol. 19, no. 2, pp. 212–234, Apr. 2008.
[30] J. R. Hollenbeck, D. R. Ilgen, D. J. Sego, J. Hedlund, D. A. Major, and J. Phillips, "Multilevel theory of team decision making: Decision performance in teams incorporating distributed expertise," J. Appl. Psychol., vol. 80, no. 2, pp. 292–316, Apr. 1995.
[31] A. Somech, "Relationships of participative leadership with relational demography variables: A multi-level perspective," J. Organizational Behav., vol. 24, no. 8, pp. 1003–1018, Dec. 2003.
[32] J. H. Hayes, T. C. Lethbridge, and D. Port, "Evaluating individual contribution toward group software engineering projects," in Proc. 25th Int. Conf. Softw. Eng., 2003, pp. 622–627.
[33] S. M. Stein and T. L. Harper, "Creativity and innovation: Divergence and convergence in pragmatic dialogical planning," J. Planning Educ. Res., vol. 32, no. 1, pp. 5–17, Sep. 2011.
[34] I. Seeber, "How do facilitation interventions foster learning? The role of evaluation and coordination as causal mediators in idea convergence," Comput. Hum. Behav., vol. 94, pp. 176–189, May 2019.
[35] X. Cheng et al., "Idea convergence quality in open innovation crowdsourcing: A cognitive load perspective," J. Manage. Inf. Syst., vol. 37, no. 2, pp. 349–376, Jun. 2020.
[36] I. Seeber, G.-J. de Vreede, R. Maier, and B. Weber, "Beyond brainstorming: Exploring convergence in teams," J. Manage. Inf. Syst., vol. 34, no. 4, pp. 939–969, Jan. 2018.
[37] E. F. Rietzschel, B. A. Nijstad, and W. Stroebe, "The selection of creative ideas after individual idea generation: Choosing between creativity and impact," Brit. J. Psychol., vol. 101, no. 1, pp. 47–68, Feb. 2010.
[38] S. H. Cady and J. Valentine, "Team innovation and perceptions of consideration: What difference does diversity make?" Small Group Res., vol. 30, no. 6, pp. 730–750, Dec. 1999.
[39] J. Wang, G. H. Cheng, T. Chen, and K. Leung, "Team creativity/innovation in culturally diverse teams: A meta-analysis," J. Organizational Behav., vol. 40, no. 6, pp. 693–708, Jul. 2019.
[40] H. Chen, P. Hsu, R. Orwig, L. Hoopes, and J. F. Nunamaker, "Automatic concept classification of text from electronic meetings," Commun. ACM, vol. 37, no. 10, pp. 56–73, Oct. 1994.
[41] B. Ramesh and A. Tiwana, "Supporting collaborative process knowledge management in new product development teams," Decis. Support Syst., vol. 27, nos. 1–2, pp. 213–235, Nov. 1999.
[42] S. Chen, L. Lin, and X. Yuan, "Social media visual analytics," Comput. Graph. Forum, vol. 36, no. 3, pp. 563–587, Jun. 2017.
[43] Y. Wu, N. Cao, D. Gotz, Y.-P. Tan, and D. A. Keim, "A survey on visual analytics of social media data," IEEE Trans. Multimedia, vol. 18, no. 11, pp. 2135–2148, Nov. 2016.
[44] P. Xu et al., "Visual analysis of topic competition on social media," IEEE Trans. Vis. Comput. Graph., vol. 19, no. 12, pp. 2012–2021, Dec. 2013.
[45] A. Marcus, M. S. Bernstein, O. Badar, D. R. Karger, S. Madden, and R. C. Miller, "Twitinfo: Aggregating and visualizing microblogs for event exploration," in Proc. SIGCHI Conf. Hum. Factors Comput. Syst., May 2011, pp. 227–236.
[46] M. Dork, D. Gruen, C. Williamson, and S. Carpendale, "A visual backchannel for large-scale events," IEEE Trans. Vis. Comput. Graphics, vol. 16, no. 6, pp. 1129–1138, Nov. 2010.
[47] K. Kucher, C. Paradis, and A. Kerren, "The state of the art in sentiment visualization," Comput. Graph. Forum, vol. 37, no. 1, pp. 71–96, Feb. 2018.
[48] N. Cao, L. Lu, Y.-R. Lin, F. Wang, and Z. Wen, "SocialHelix: Visual analysis of sentiment divergence in social media," J. Visualizat., vol. 18, no. 2, pp. 221–235, May 2015.
[49] L. Zhang, H. Yuan, and R. Y. K. Lau, "Predicting and visualizing consumer sentiments in online social media," in Proc. IEEE 13th Int. Conf. e-Bus. Eng. (ICEBE), Nov. 2016, pp. 92–99.

[50] S. Liu, J. Yin, X. Wang, W. Cui, K. Cao, and J. Pei, "Online visual analytics of text streams," IEEE Trans. Vis. Comput. Graphics, vol. 22, no. 11, pp. 2451–2466, Dec. 2016.
[51] D. J. Watts and S. H. Strogatz, "Collective dynamics of 'small-world' networks," Nature, vol. 393, no. 6684, pp. 440–442, 1998.
[52] Z. S. Harris, "Distributional structure," Word, vol. 10, nos. 2–3, pp. 146–162, 1954.
[53] D. M. Blei, A. Y. Ng, and M. I. Jordan, "Latent Dirichlet allocation," J. Mach. Learn. Res., vol. 3, pp. 993–1022, Mar. 2003.
[54] T. Mikolov, I. Sutskever, K. Chen, G. Corrado, and J. Dean, "Distributed representations of phrases and their compositionality," in Proc. Adv. Neural Inf. Process. Syst., vol. 26, 2013, pp. 1–9.
[55] Q. Le and T. Mikolov, "Distributed representations of sentences and documents," in Proc. ICML, 2014, pp. 1188–1196.
[56] J. H. Lau and T. Baldwin, "An empirical evaluation of doc2vec with practical insights into document embedding generation," in Proc. 1st Workshop Represent. Learn. (NLP), 2016, pp. 78–86.
[57] I. T. Jolliffe and J. Cadima, "Principal component analysis: A review and recent developments," Phil. Trans. R. Soc. A, vol. 374, Apr. 2016, Art. no. 20150202.
[58] A. Mead, "Review of the development of multidimensional scaling methods," J. Roy. Stat. Soc., D Statistician, vol. 41, no. 1, pp. 27–39, 1992.
[59] L. van der Maaten and G. Hinton, "Visualizing data using t-SNE," J. Mach. Learn. Res., vol. 9, pp. 2579–2605, Nov. 2008.
[60] C. Viau, M. J. McGuffin, Y. Chiricota, and I. Jurisica, "The FlowVizMenu and parallel scatterplot matrix: Hybrid multidimensional visualizations for network exploration," IEEE Trans. Vis. Comput. Graphics, vol. 16, no. 6, pp. 1100–1108, Nov. 2010.
[61] T. Buering, J. Gerken, and H. Reiterer, "User interaction with scatterplots on small screens—A comparative evaluation of geometric-semantic zoom and fisheye distortion," IEEE Trans. Vis. Comput. Graphics, vol. 12, no. 5, pp. 829–836, Sep. 2006.
[62] N. Elmqvist and J. Fekete, "Hierarchical aggregation for information visualization: Overview, techniques, and design guidelines," IEEE Trans. Visualizat. Comput. Graphics, vol. 16, no. 3, pp. 1100–1108, Jun. 2010.
[63] S. Chen, S. Chen, L. Lin, X. Yuan, J. Liang, and X. Zhang, "E-Map: A visual analytics approach for exploring significant event evolutions in social media," in Proc. EEE Conf. Vis. Anal. Sci. Technol., Oct. 2017, pp. 36–47.
[64] M. Choi et al., "TopicOnTiles: Tile-based spatio-temporal event analytics via exclusive topic modeling on social media," in Proc. CHI Conf. Hum. Factors Comput. Syst., Apr. 2018, pp. 1–11.
[65] M. A. Syakur, B. K. Khotimah, E. M. S. Rochman, and B. D. Satoto, "Integration K-means clustering method and elbow method for identification of the best customer profile cluster," IOP Conf. Ser., Mater. Sci. Eng., vol. 336, Nov. 2017, Art. no. 012017.
[66] D. Auber, Y. Chiricota, F. Jourdan, and G. Melancon, "Multiscale visualization of small world networks," in Proc. IEEE Symp. Inf. Visualizat., Oct. 2003, pp. 75–81.
[67] B. Teimourpour and B. Asgharpour, "Improving circular layout algorithm for social network visualization using genetic algorithm," Appl. Data Manage. Analysis., pp. 17–28, Oct. 2018.
[68] S. Cao et al., "An agent-based model of leader emergence and leadership perception within a collective," Complexity, vol. 2020, pp. 1–11, Apr. 2020.